\documentclass[a4paper]{jpconf}
\usepackage{graphicx}
\usepackage{iopams}
\begin{document}
\title{Parallelized Kalman-Filter-Based Reconstruction of Particle Tracks on Many-Core Architectures}

\author{G Cerati$^4$,
P Elmer$^2$, S Krutelyov$^1$, S Lantz$^3$, M Lefebvre$^2$, M Masciovecchio$^1$, K McDermott$^3$, D Riley$^3$,
M Tadel$^1$, P Wittich$^3$, F Würthwein$^1$, and A Yagil$^1$}

\address{$^1$ UC San Diego, 9500 Gilman Dr., La Jolla, California, USA 92093}
\address{$^2$ Princeton University, Princeton, New Jersey, USA 08544}
\address{$^3$ Cornell University, Ithaca, New York, USA}
\address{$^4$ Fermi National Accelerator Laboratory, Batavia, USA}

\ead{Daniel.Riley@cornell.edu}

\begin{abstract}
Faced with physical and energy density limitations on clock speed, contemporary microprocessor designers have increasingly turned to on-chip parallelism for performance gains. Algorithms should accordingly be designed with ample amounts of fine-grained parallelism if they are to realize the full performance of the hardware. This requirement can be challenging for algorithms that are naturally expressed as a sequence of small-matrix operations, such as the Kalman filter methods widely in use in high-energy physics experiments. In the High-Luminosity Large Hadron Collider (HL-LHC), for example, one of the dominant computational problems is expected to be finding and fitting charged-particle tracks during event reconstruction; today, the most common track-finding methods are those based on the Kalman filter. Experience at the LHC, both in the trigger and offline, has shown that these methods are robust and provide high physics performance. Previously we reported the significant parallel speedups that resulted from our efforts to adapt Kalman-filter-based tracking to many-core architectures such as Intel Xeon Phi. Here we report on how effectively those techniques can be applied to more realistic detector configurations and event complexity.
\end{abstract}
\section{Introduction}

\begin{figure}[htb]
  \begin{minipage}[c]{0.55\textwidth}
    \vspace{0pt}
    \includegraphics[width=\textwidth]{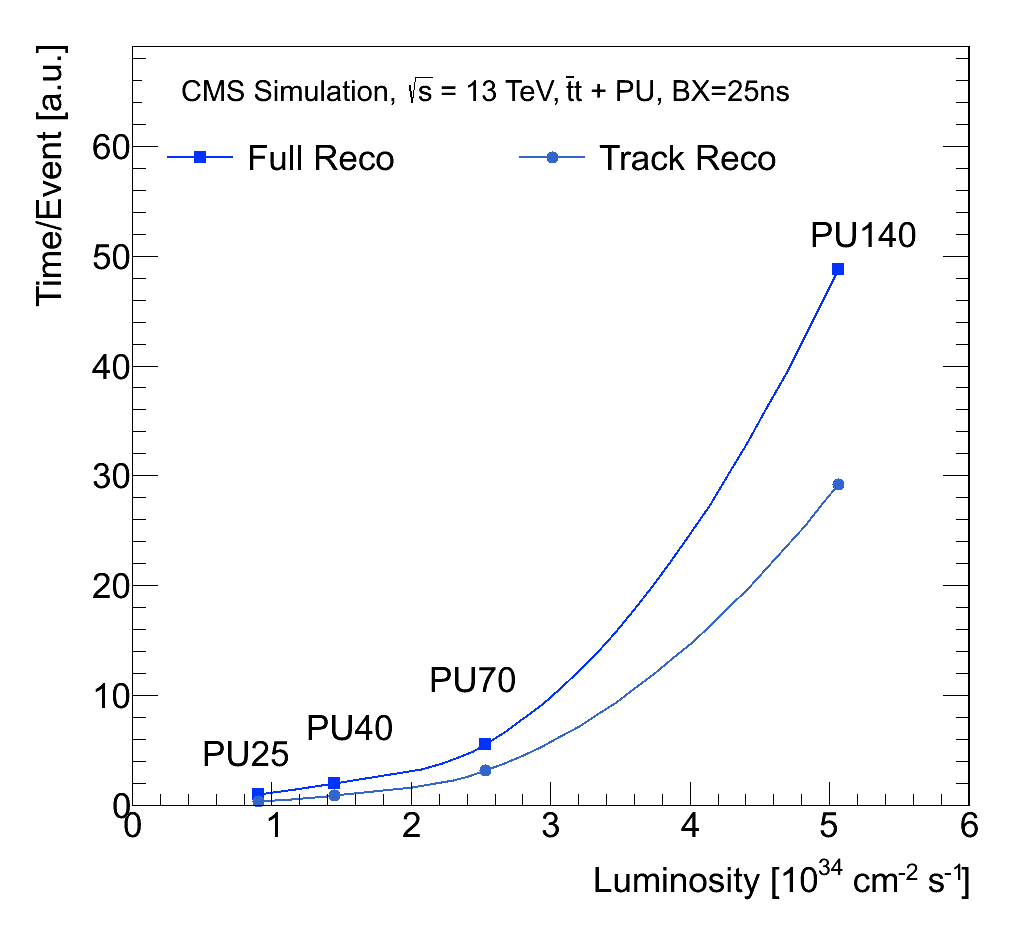}
  \end{minipage}
  \hspace*{\fill}
  \begin{minipage}[c]{0.35\textwidth}
    \vspace{0pt}
    \caption{\raggedright CPU time per event versus instantaneous luminosity, for both full reconstruction and the dominant tracking portion. Simulated data with pile-up of 25 primary interactions per event (PU25) corresponds to the data taken during 2012, while pile-up of 140 (PU140) corresponds to the low end of estimates for the HL-LHC era.}
    \label{fig:eff_tracking_pileup}
  \end{minipage}
  \hspace*{\fill}
\end{figure}

Finding tracks is the most computationally complex of the steps reconstructing events in the CMS detector. The speed of online reconstruction has a direct impact on how effectively interesting data can be selected from the 40 MHz collisions rate, while the speed of the offline reconstruction limits how much data can be processed for physics analyses.  The time spent in tracking will become even more important in the HL-LHC era of the Large Hadron Collider, as the increase in event rate will increase the detector occupancy (“pile-up”, PU), leading to the exponential increases in track reconstruction time, illustrated in Figure~\ref{fig:eff_tracking_pileup}~\cite{pileup}.

At the same time, due to power density and physical scaling limits, the performance of single CPUs has slowed, with advances in performance increasingly relying on highly parallel architectures.  In order to sustain the higher HL-LHC processing requirements without compromising physics performance or unduly delaying results, the LHC experiments must make full use of highly parallel architectures.

As a potential solution, we are investigating adapting the existing CMS track finding algorithms and data structures~\cite{cmstrack} to make efficient use of highly parallel architectures, such as Intel’s Xeon Phi and NVIDIA general-purpose graphics processing units (GPGPUs).  In this investigation we have followed a staged approach, starting with Intel Xeon and Xeon Phi Knights Corner (KNC) architectures, an idealized detector geometry, and a series of simpler ``warm-up'' exercises such as track fitting.  This simplified problem domain was used to develop our tools, techniques, and understanding of the issues scaling track finding to many cores.  The warm-up exercises let us develop useful components while also allowing the physicists to become familiar with the computational tools and techniques, while the computational experts learned about the problem domain.  Armed with the results of those initial investigations, we are now addressing more realistic detector geometries and event content, as well as adding new platforms.  This paper gives an overview of our progress to date and assesses the effectiveness of our staged approach.

\section{Kalman Filter Tracking}

Our targets for parallel processing are track reconstruction and fitting algorithms based on the Kalman Filter~\cite{kalman} (KF). KF-based tracking algorithms are widely used to incorporate estimates of multiple scattering directly into the trajectory of the particle. Other algorithms, such as Hough Transforms and Cellular Automata~\cite{cellular2}\cite{cellular3}, are more naturally parallelized.  However, these are not the main algorithms in use at the LHC today. The LHC experiments have an extensive understanding of the physics performance of KF algorithms; they have proven to be robust and perform well in the difficult experimental environment of the LHC.

KF tracking proceeds in three main stages: seeding, building, and fitting. Seeding provides the initial estimate of the track parameters based on a few hits in the innermost regions of the detector; seeding is currently out of scope for our project. Track building projects the track candidate outwards to collect additional hits, using the KF to estimate which hits represent the most likely continuation of the track candidate. Track building is the most time consuming step, as it requires branching to explore multiple candidate tracks per seed after finding compatible hits on a given layer. When a complete track has been reconstructed, a final fit using the KF is performed to provide the best estimate of the track's parameters.

To take full advantage of parallel architectures, we need to exploit two types of parallelism: vectorization and parallelization. Vector operations perform a single instruction on multiple data (SIMD) at the same time, in lockstep. In tracking, branching to explore multiple candidates per seed can interfere with lock-step single-instruction synchronization. Multi-thread parallelization aims to perform different instructions at the same time on different data. The challenge to track building is workload balancing across different threads, as track occupancy in a detector is not uniformly distributed on a per event basis. Past work by our group has shown progress in porting sub-stages of KF tracking to support parallelism in simplified detectors~\cite{cerati-chep15}\cite{ctd-2016}\cite{chep-2016}\cite{ctd-2017}. As the hit collection is completely determined after track building, track fitting can repeatedly apply the KF algorithm without branching, making this a simpler case for both vectorization and parallelization; results in implementing vectorized parallel KF tracking to Xeon Phi were shown previously~\cite{acat2014}.

\subsection{Matriplex}

The implementation of a KF-based tracking algorithm consists of a sequence of operations on matrices of sizes from $3\times 3$ up to $6\times 6$.  In order to optimize efficient vector operations on small matrices, and to decouple the computational details from the high level algorithm, we have developed a new matrix library. The {\it Matriplex} memory layout uses a matrix-major representation optimized for loading vector registers for SIMD operations on small matrices, using the native vector-unit width on processors with small vector units or very large vector widths on GPGPUs.  Matriplex includes a code generator for defining optimized matrix operations, with support for symmetric matrices and on-the-fly matrix transposition. Patterns of elements that are known by construction to be zero or one can be specified, and the resulting  code will be optimized to eliminate unnecessary register loads and arithmetic operations. The generated code can be either standard C++ or macros that map to architecture-specific intrinsic functions.

\section{Initial Test Scenario}\label{sec:scenarios}

In order to study parallelization with minimal distractions, we developed a standalone KF-based tracking code using a simplified ideal barrel geometry with a uniform longitudinal magnetic field, gaussian-smeared hit positions, a particle gun simulation with flat transverse momentum distribution between 0.5 and 10 GeV, no material interaction, and no correlation between particles nor decays.  This simplified configuration was used to study vector and parallel performance and to study the performance of different choices of coordinate system and handling of KF-updates.  These studies led to the use of a hybrid coordinate system, using global Cartesian coordinates for spatial positions and polar coordinates for the momentum vector.

\subsection{Initial Results}

\begin{figure}[htb]
  \begin{minipage}{18pc}
   \includegraphics[width=18pc]{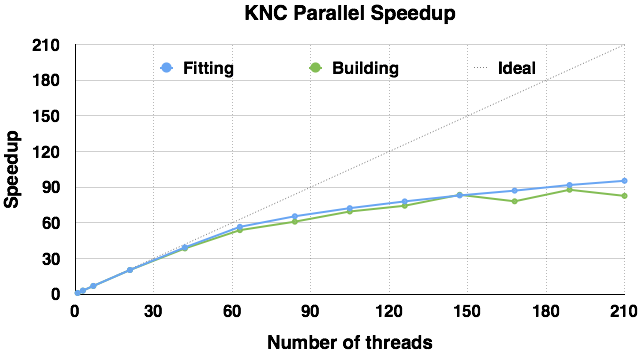}
  \end{minipage}\hspace{2pc}%
  \begin{minipage}{18pc}
   \includegraphics[width=18pc]{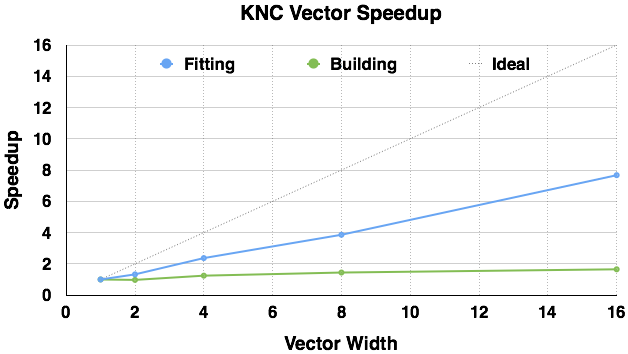}
  \end{minipage}
 \caption{KNC parallel and vector speedups for track fitting and building.}
 \label{fig:speedup}
\end{figure}

Figure~\ref{fig:speedup} illustrates the parallel and vector speedups on the KNC platform for the simple test scenario.  We achieve good parallelization up to the number of physical cores, 61, with some continued improvements from hyper-threading (KNC CPUs change hyper-thread every clock cycle, so full utilization would be a speedup of 122).  While both track fitting and building are ``embarrassingly parallel'', we believe the parallelization speedup is limited by memory bandwidth and cache sizes, so our scaling performance is somewhat dependent on the details of the memory subsystem.  For the relatively simple track fitting we also see good vector performance.  The combinatorial nature of the track building algorithm, which examines and adds a variable number of hit candidates in each layer, results in many branching operations that impede vectorization, as well as adding frequent repacking operations to keep the full vector width utilized.  The larger and more complicated data structures used for selecting hit candidates also results in poorer data locality and higher bandwidth requirements.  We are continuing to investigate strategies for improving track building vector utilization.

\subsection{Initial Lessons Learned}

Achieving acceptable vector and multi-thread parallel performance required careful attention to detail.  Regular profiling with Intel VTune Amplifier and examination of the compiler optimization reports helped identify many obstacles, some relatively subtle.  For multi-thread parallelism, the two most critical issues were memory management and workload balancing.  To avoid memory stalls and cache conflicts, we reduced our data structures to the minimum necessary for the algorithm, optimized our data structures for efficient vector operations, and minimized object instantiations and dynamic memory allocations.  Workload balancing for track building is complicated by the uncertain distribution of track candidates and hits, which can result in large tail effects for a naive static partitioning scheme.  We found that using Intel Threaded Building Blocks~\cite{tbb} (TBB) tasks, with smaller units of work and dynamic ``work-stealing'', let us naturally express thread-parallelism in a way that allowed more dynamic allocation of resources and reduced the tail effects from uneven workloads.

\section{Adding Complications}

Recently we have nearly completed work on using more realistic detector geometries and realistic events.  
In addition, we have added new platforms, principally the NVIDIA Tesla K20/K40 and Pascal P100 GPGPUs, which present a very different programming model.

\subsection{Realistic Geometries}

\begin{figure}[htb]
  \begin{minipage}{16pc}
   \includegraphics[width=16pc]{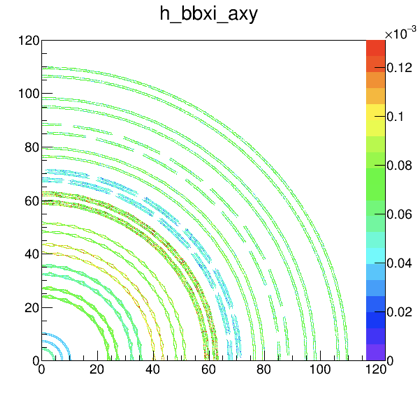}
  \end{minipage}\hspace{2pc}%
  \begin{minipage}{16pc}
   \includegraphics[width=16pc]{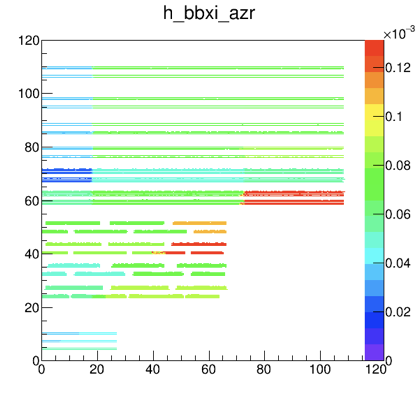}
  \end{minipage}
  \caption{x-y and r-z views of simplified CMS geometry}
  \label{fig:cms-geom}
\end{figure}

We implemented two more realistic geometries.  The first adds endcaps to our idealized geometry, yielding a ``cylinder with lids'' that was used for initial algorithm development.  The second geometry implements a simplified form of the CMS tracker geometry, illustrated in Figure~\ref{fig:cms-geom}.  For track building in the less regular geometry, track propagation is performed in two steps, an initial propagation to the average radius of the barrel or average Z of the endcap disk, followed by further propagation steps to the position of each candidate hit.  KF updates are performed on the plane tangent to the hit radius.  These choices avoid the complexities of the full detector description.  We are currently validating the physics performance of our algorithms with the simplified CMS geometry, prior to completing computational performance studies.

\subsection{Realistic Events}

\begin{figure}[htb]
  \hspace*{\fill}
  \begin{minipage}[c]{0.55\textwidth}
    \vspace{0pt}
    \includegraphics[width=\textwidth]{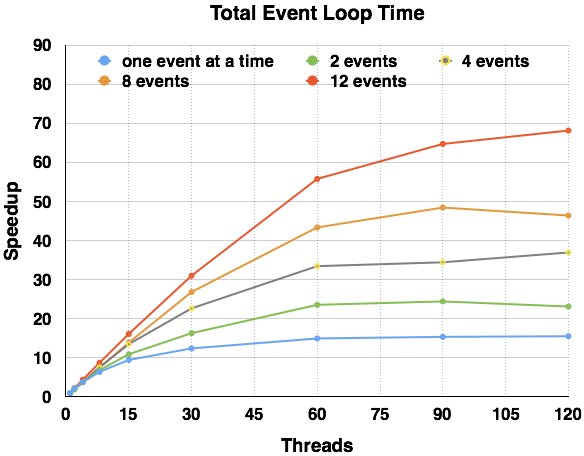}
  \end{minipage}
  \hspace*{\fill}
  \begin{minipage}[c]{0.35\textwidth}
    \vspace{0pt}
    \caption{\raggedright KNC track building speedup as a function of the number of threads, for events with occupancy similar to seeds from the first iteration of the CMS iterative tracker.  For these sparser events, processing tracks from multiple events in parallel significantly improves scaling.}
    \label{fig:multi-events}
  \end{minipage}
  \hspace*{\fill}
\end{figure}

Event data from the full CMS simulation suite is translated into a reduced form that can be processed by our standalone tracking code. Seeds used are from the first iteration of the CMS iterative tracker~\cite{pileup}.  Realistic events typically have lower occupancy and less uniformity in track distribution, leading to more challenges for workload balancing and thus to worse multi-thread scaling than our ``toy'' setup.  To counter this effect we process multiple events at a time, filling in gaps caused by the varying levels of parallelism within a single event.  Figure~\ref{fig:multi-events} shows that this strategy is effective at recovering good parallel scaling for more realistic events, overcoming the workload balancing issues from lower occupancy events.

\subsection{GPGPU Developments}

\begin{figure}[htb]
  \begin{minipage}[c]{0.5\textwidth}
    \vspace{0pt}
    \includegraphics[width=\textwidth]{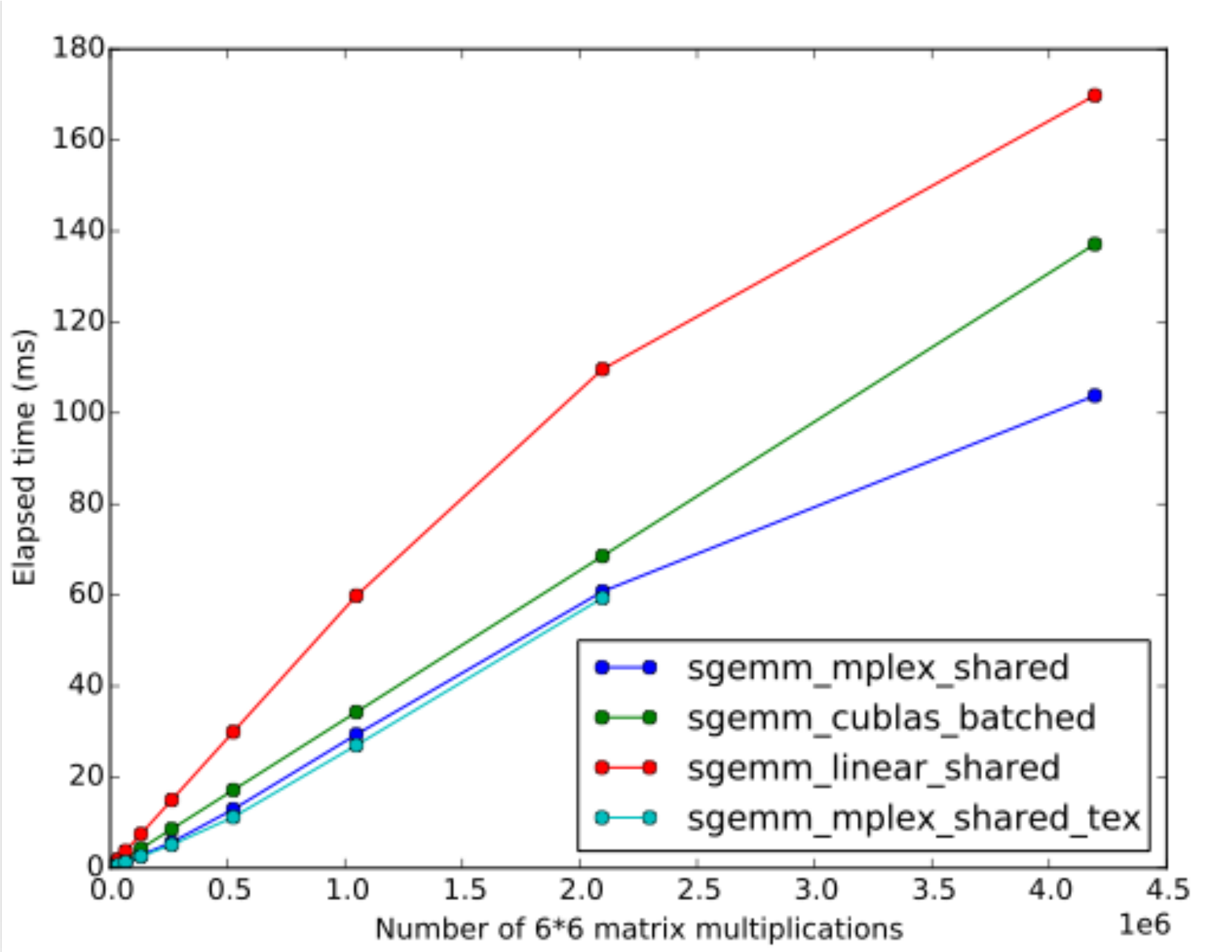}
  \end{minipage}
  \hspace*{\fill}
  \begin{minipage}[c]{0.4\textwidth}
    \vspace{0pt}
    \caption{\raggedright Performance comparison of different memory organizations for multiplying $6 \times 6$ matrices, showing the time as a function of the number of $6 \times 6$ matrix multiplications.  Compared to arrays of matrices using standard libraries, Matriplex (labeled ``mplex'') shows superior performance.}
    \label{fig:gpu-ds}
  \end{minipage}
  \hspace*{\fill}
\end{figure}

We have also implemented the track fitting and building algorithms on NVIDIA GPGPUs using the CUDA toolkit.  The large core count of GPGPUs and their thread scheduling policies force a large number of threads to be concurrently scheduled.  This implies that coping with branching is critical as threads of a {\it warp} (a group of 32 consecutive threads) execute the same common instructions at the same time.  Investigating the memory layout efficiency for GPGPUs, we found Matriplex to be the most efficient of the alternatives we tested as it naturally allows coalesced memory accesses, where memory requests from consecutive threads to consecutive locations can be coalesced into a single transaction.  Figure~\ref{fig:gpu-ds} shows the results of these studies.  Using the Matriplex data structure also facilitated code sharing, by templating low level routines such as KF propagation to accept Matriplex structures sized for CPU or GPGPU, while only high-level ``steering'' routines needed to be customized.  To minimize branching, the GPGPU version allocates a fixed number of slots for each track seed to store the corresponding track candidates from that seed.  The candidates are organized as a heap, allowing rapid identification of the best candidates.  As in the CPU version, we found that multiple events in flight was also necessary for scaling so that expensive copy operations could be overlapped with processing.  We are currently investigating whether the design choices made for the GPGPU implementation can be applied to improve our vector performance on the Xeon platforms.

\section{Conclusion and Outlook}

We have made significant progress in parallelized and vectorized Kalman Filter-based end-to-end tracking R\&D on Xeon and Xeon Phi architectures, with some  work on GPGPUs. Through the use of a variety of tools we have developed a good understanding of bottlenecks and limitations of our implementation which has led to further improvements. With our own Matriplex package and TBB, we can achieve good utilization of unconventional highly parallel vector architectures; these results should also be applicable to newer architectures with large vector registers, such as Intel Skylake-SP.  We are currently focusing on  processing fully realistic data.

We found our staged approach to the problem to be effective in developing our understanding of the problem and promoting effective collaboration across domains.  Despite some initial concerns about diluting our efforts, we have also found that addressing multiple architectures has been useful for developing approaches that can be applied across architectures.

\ack

This work is supported by the U.S. National Science Foundation, under the grants PHY-1520969, PHY-1521042, PHY-1520942 and PHY-1120138, and by the U.S. Department of Energy.

\end{document}